# Size-Distribution Scaling in Clusters of Allelomimetic Agents

Dranreb Earl Juanico and Caesar Saloma

National Institute of Physics
University of the Philippines, Diliman
dojuanico@up.edu.ph

## Extended Abstract

This article presents a simple, yet general, model that explains the scaling properties observed in real cluster-size distributions.

The model defines a threshold response for an agent expressed by

$$P(\sigma, \beta) = \begin{cases} \beta, & \text{if } \sigma > \sigma_c \\ 0, & \text{if } \sigma \le \sigma_c \end{cases} \quad (1)$$

Equation (1) gives the probability that an agent actuates depending on its stimulus level $\sigma$ and $\sigma_c$ is some threshold value. Such characteristic is akin to neuronal stimulation. Furthermore, equation (1) states that an agent may override the consequence of $\sigma > \sigma_c$ by a complementary "choice" of not performing an action. This is reflected by $\beta$ which may be less than 1. Equation (1) therefore incorporates the ability of an agent to decide. Agents are distributed in a $d$-dimensional lattice of length $L$. A fraction $p$ of the agent population is designated to be *unresponsive*. Unresponsive agents are characterized by $\beta=0$ such that $P(\sigma,\beta)=0$ even if $\sigma>\sigma_c$. These agents may be thought of as impurities in the lattice. The value of $p$ is varied between 0 and 1.

The lattice is constantly bombarded by external stimuli (e.g., environmental conditions or public information in the form of advertisements, and the like). The consequence of this is that at each time step of the numerical experiment, $\sigma \rightarrow \sigma+1$ for all agents (i.e., stimulus levels are raised by a unit).

An agent is then randomly chosen, say an agent at cell i,j in a $d=2$ lattice, to behave according to (1) with $\beta=1$. Hence, if $\sigma(i,j)$ exceeds $\sigma_c(i,j)$ then this agent outputs a particular action **A**. Allow us to distinguish this chosen agent as a *harbinger*. The harbinger is the initiator of an action. By the performance of **A** the harbinger's stimulus level decreases: $\sigma(i,j) \rightarrow \sigma(i,j)-2d$, as though releasing tension. Meanwhile, the stimulus level of each of the harbinger's $2d$ nearest-neighbors is increased by a unit due to their observance of **A**. For a responsive neighbor, sufficient stimulation ($>\sigma_c$) resulting from the observance of **A** makes it actuate **A** with a probability $\beta$. Let us assume that $\beta$ is of a particular value $\alpha$, wherein $\alpha$ is defined as the *allelomimesis index*. In contrast, any amount of stimulation brought about by the observance of **A** will have no effect on an unresponsive neighbor.

The harbinger's neighbors in turn pass around the information to their corresponding neighbors by actuating **A**. The dynamics of action propagation can be summarized as two fundamental processes: (i) the selection of a harbinger that initiates an action, and (ii) propagation of action through nearest-neighbor connections. Process (ii) is repeated until the action initiated by the harbinger ceases to propagate. All agents that actuate or have actuated **A** are considered to belong to a cluster and the total number of these agents corresponds to the size $s$ of the cluster. The resulting distribution $f(s)$ is highly-dependent on $\alpha$ and $p$.

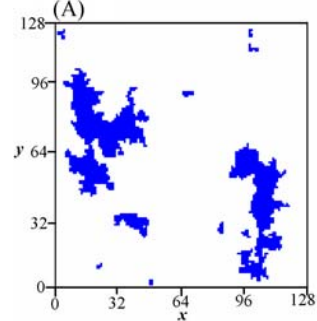

**Figure 1 Clusters that result for $\alpha=1$ and $p=0$**

Figure 1 illustrates the morphology of clusters resulting from the model. These clusters apparently exhibit fractal structures that were produced by models of urban growth through *diffusion-limited aggregation* [1].

The model is able to fit cluster-size distributions of actual clustering systems in nature. This is done by simply tuning the values of $\alpha$ and $p$. Our significant finding is that $\alpha$ have high-values for animals implying that allelomimesis is strongly expressed whereas $\alpha \sim 0.3$ for human beings. Details of the data analysis method utilized in relating the model with data gathered from multifarious real clustering is discussed in Ref. 2.

# Size-Distribution Scaling in Clusters of Allelomimetic Agents

Dranreb Earl Juanico and Caesar Saloma

National Institute of Physics
University of the Philippines, Diliman
dojuanico@up.edu.ph

## Abstract

*The allelomimesis clustering model is based on only two parameters: a local parameter $\alpha$ that represents the probability of nearest-neighbor copying and a global parameter p that represents the fraction of unresponsive agents. The model results into the formation of clusters of agents, the sizes of which obey a distribution that is determined by the values of $\alpha$ and p. Several experimental data are fitted by tuning the two parameters. In particular, the significance of the value of $\alpha$ that corresponds to an experimental data is discussed and is justified according to behavioral context. Recommendations for possible extensions of the model are also enumerated.*

## 1. Introduction

The penultimate hallmark of complex systems is the principle of emergence – macroscopic regularity of the system arising from apparently irregular (disordered) microscopic interactions between the system's constituents. Among the most commonly observed of these emergent properties is *clustering*.

Clusters in nature exist in different sizes. An interesting and somewhat unexpected observation is the regularity of the statistical distribution of cluster sizes, which generally obeys what is known as a *power-law*. If we denote by $s$ the size of a cluster and by $f(s)$ the frequency of occurrence of $s$ (i.e., the number of times that a particular value of $s$ is tallied), then one would witness that the plot of $f(s)$ versus $s$ in double logarithmic scale is a decreasing straight line. The slope of this line corresponds to the exponent of the power law. Let us refer to the value of this slope as $\tau$.

Power-law behavior is a well-known result in the area of complex systems. This article presents a simple, yet general, mechanism that leads to this power-law behavior.

In section 2, a phenomenological model based on allelomimetic behavior is discussed. Allelomimesis is the tendency of individuals to imitate its neighbors; hence, allelomimetic behavior is the best candidate as local interaction that could lead to the formation of clusters. The results of the model are discussed in Section 3. A comparison of these results with several cluster systems found in nature is also presented. A few recommendations for further extension of this study are pointed out in Section 4.

## 2. The Clustering Model

### 2.1 Behavior of a single agent

The probability that an agent "performs" a certain "action" is described as follows:

$$P(\sigma) = \begin{cases} 1, & \text{if } \sigma > \sigma_c \\ 0, & \text{if } \sigma \leq \sigma_c \end{cases} \quad (1)$$

wherein $\sigma$ is the total stimulus received by an agent (both from its environment and its neighboring agents) and $\sigma_c$ is an arbitrary threshold stimulus level. Equation (1) means that an agent actuates if its stimulus level exceeds the threshold. Such characteristic is akin to neuronal stimulation [1].

In reality, however, an agent may override the consequence of $\sigma > \sigma_c$ by a complementary "choice" of not performing an action. Let us assume that the probability the agent will not actuate if $\sigma > \sigma_c$ is $1-\beta$. Hence, (1) may be rewritten in the following form:

$$P(\sigma, \beta) = \begin{cases} \beta, & \text{if } \sigma > \sigma_c \\ 0, & \text{if } \sigma \leq \sigma_c \end{cases} \quad (2)$$

Equation (2) incorporates the ability of an agent to decide.

Generally, $\beta$ and $\sigma_c$ may vary among agents. But to make the model as simple as possible, it is assumed that $\beta$ is a mean value over a population of agents, hence, $\beta$ is a constant with respect to a particular agent population. However, $\beta$ is allowed to vary between different populations. On the other hand, two cases are considered in assigning the value of $\sigma_c$ – it is either fixed ($\sigma_c = 4$) for all agents or it varies within a range ($2 \leq \sigma_c \leq 16$) among agents.

### 2.2 Lattice of agents

Agents are distributed in a *d*-dimensional lattice of length $L$ consisting of $L^d$ discrete cells. A cell may only accommodate a single agent. Thus, $L^d$ also corresponds to the size of the agent population.

A fraction *p* of the agent population is designated to be *unresponsive*. Unresponsive agents are characterized by β=0 such that $P(\sigma,\beta)=0$ even if $\sigma>\sigma_c$. These agents may be thought of as impurities in the lattice. The value of *p* is varied between 0 and 1.

## 2.3 Dynamics of action propagation

The lattice is constantly bombarded by external stimuli (e.g., environmental conditions or public information in the form of advertisements, and the like). The consequence of this is that at each time step of the numerical experiment, $\sigma \to \sigma+1$ for all agents (i.e., stimulus levels are raised by a unit).

An agent is then randomly chosen, say an agent at cell i,j in a *d*=2 lattice, to behave according to (1). Hence, if σ(i,j) exceeds $\sigma_c$(i,j) then this agent outputs a particular action **A**. Otherwise, nothing happens and another agent is randomly chosen. Allow us to distinguish this chosen agent as a *harbinger*. The harbinger is the initiator of an action. Once a harbinger is selected, the bombardment of external stimuli is momentarily paused to allow us to focus on the consequence of the harbinger's action to the entire lattice. Furthermore, we assume that there can only be one harbinger at a time but any agent is a potential harbinger at any given time.

By the performance of **A** the harbinger's stimulus level decreases: σ(i,j)→σ(i,j)−2*d*, as though releasing tension. Meanwhile, the stimulus level of each of the harbinger's 2*d* nearest-neighbors is increased by a unit due to their observance of **A**. These neighbors behave according to (2) to decide whether or not to mimic the harbinger and actuate **A**. For a responsive neighbor, sufficient stimulation (>$\sigma_c$) resulting from the observance of **A** makes it actuate **A** with a probability β. Let us assume that β is of a particular value α. In contrast, any amount of stimulation brought about by the observance of **A** will have no effect on an unresponsive neighbor.

The harbinger's neighbors in turn pass around the information to their corresponding neighbors by actuating **A**. This propagation continues up to the last agent that performs **A** without influencing its corresponding nearest neighbors.

The parameter α is defined as the *allelomimesis index*. Its value is tuned between 0 and 1. On one hand, α=0 is equivalent to setting *p*=1, i.e. all agents in the population are unresponsive to their neighbors; hence, non-copying or non-allelomimetic. On the other hand, α=1 implies a highly-allelomimetic population of agents wherein imitation of neighbors is a big factor that promotes clustering.

The dynamics of action propagation can be summarized as two fundamental processes: (i) the selection of a harbinger that initiates an action, and (ii) propagation of action through nearest-neighbor connections.

## 2.4 Cluster and cluster size

Process (ii) is repeated until the action initiated by the harbinger ceases to propagate. All agents that actuate or have actuated **A** are considered to belong to a cluster and the total number of these agents corresponds to the size of the cluster *s*. Subsequently, the bombardment of external stimuli is resumed for the proceeding time step. Process (i) results to the initiation of another action and process (ii) propagates this action through the lattice, hence, establishing the formation of another cluster. By repeating processes (i) and (ii) over several time steps, one generates different values of *s*. This allows one to deduce the statistical distribution *f*(*s*).

## 3. Results and Discussion

### 3.1 Numerical simulations

Figure 1 illustrates the morphology of clusters at different settings of the parameters α and *p*. These clusters apparently exhibit fractal structures resembling those that were produced by models of urban growth through *diffusion-limited aggregation* [2].

Let us first deal with the effect of varying α by setting *p* = 0. Figure 2 plots the power-law cluster size distribution (CSD) in double logarithmic scale for different values of α. Notice how the lines steepen with increasing value of α, indicating that the scaling exponent τ is negatively correlated with the parameter α. It is expected that τ→∞ as α→0, consistent with a dirac-delta CSD centered at *s*=1 for α=0 (i.e., no clusters are formed).

To show the effect of the parameter *p* on the CSD, we fix α to a value of 1. Figure 3 exhibits a distortion of the CSD at large values of *s*. The degree of such distortion intensifies with increasing *p*.

Considering that thresholds $\sigma_c$ may vary among agents, we compare the set of CSDs (with different α and *p* = 0) for the case wherein 2≤$\sigma_c$ ≤ 16 with the set for which $\sigma_c$=4. Figure 4 plots the CSDs as data points in the former case and as broken lines in the latter case. There is no observable difference and this implies that the exact value of $\sigma_c$ of an agent does not affect the CSD. Hence, the CSD is robust to variations of $\sigma_c$ within an agent population.

## 3.2 Comparison with data for real systems

We fit our model to different CSDs taken from experimental observations of actual clusters. The goodness-of-fit is measured in terms of the mean square error (MSE) between the data and the curve generated from the model. Figures 5 and 6 present data on four animal systems and four distinct human cluster systems, respectively.

Remarkably, $\alpha$ is high for animal systems (except for Serengeti lions) implying that allelomimesis is strongly expressed in animals. According to Wagner and Danchin, "conspecific copying" (or allelomimesis) is a ubiquitous mechanism behind the formation of aggregates such as leks and colonies [3]. Bonabeau and Dagorn showed that "biosocial attraction" (another form of allelomimesis) promotes schooling in fishes [4]. Parrish and Keshet further proposed that allelomimesis is a generic mechanism that maintains the cluster as a cohesive unit [5]. Indeed, ecological evidence for a high value of $\alpha$ in the animal kingdom is compelling. The seemingly low $\alpha$ (=0.1) for Serengeti lions is compensated by the high-value of $p$. Such disparity may be explained by the fact that the lions that were observed by Schaller were *nomadic* [6]. This means that they are likely to wander alone or in small groups, hence, even though lions may be considered highly-social (high $p$), being nomadic disrupts the information flow between lions resulting in low $\alpha$. A comprehensive discussion of the deduced values of $\alpha$ and $p$ is found in Ref. 7.

The value of $\alpha$ for human cluster systems is low as compared to animal systems, implying that allelomimesis is only moderately expressed in human beings. This can be justified by considering that humans are generally more highly cognitive than animals, which consequently overrides their instinctive tendency to be allelomimetic. Furthermore, telecommunication technology (which only humans are capable of) diminishes the requirement of information transfer through nearest-neighbor connections such as allelomimesis. Interestingly, $\alpha$ is not significantly different among distinct human cluster systems ($\alpha \sim 0.3$). This result is quite expected because even though we consider clusters of cities, of households or of employees to be distinct from one another, one fact remains common between them – these systems are all made up of human beings. It would be worthwhile to investigate the origin of such seemingly universal value of $\alpha$ from a psychological point of view.

## 4. Recommendations

The model due to its inherent simplicity has cut down on details as much as possible so that it can be considered generic, hence, applicable to a wide variety of systems. Here, we suggest some minor points of modification to allow a more realistic description.

The stimulation on an agent due to constant bombardment of external factors may not necessarily be equal to unity (i.e., $\sigma \rightarrow \sigma + 1$). It can be expressed as $\sigma \rightarrow \sigma + \eta$, where $\eta$ represents a positive Poisson number that appropriately describes the time fluctuation of the amount of external stimuli. Furthermore, subsequent stimulation of neighboring agents may not necessarily decrease the stimulus level of the harbinger by an amount that is equal to the number of its nearest neighbors. That is, one can write $\sigma \rightarrow \sigma - \varepsilon$, where $\varepsilon$ is a positive number derived from a Gaussian or a Binomial probability distribution. It follows that the ensuing stimulation of responsive neighbors that observe the actuation of the harbinger can be expressed as $\sigma_n \rightarrow \sigma_n + \delta_n$ where the subscript $n$ represents the nearest-neighbor and $\sum_n \delta_n = \varepsilon$. Note here that $\delta_n$ could either be positive or negative, implying that the stimulation is excitatory or inhibitory, respectively [1].

## 5. Conclusion

A simple model of cluster formation is proposed to explain the cluster size distribution observed for various cluster systems in nature. The model consists of two mutually independent parameters, namely $\alpha$ and $p$. The value of $\alpha$ represents the probability that an agent mimics the action of its nearest-neighbors whereas $p$ is the fraction of unresponsive agents that characterizes the particular agent population. Resulting CSDs are highly-dependent on $\alpha$ and $p$.

The model fits into experimental data corresponding to various cluster systems in nature. High value of $\alpha$ generally characterizes animal systems whereas $\alpha \sim 0.3$ distinguishes human cluster systems.

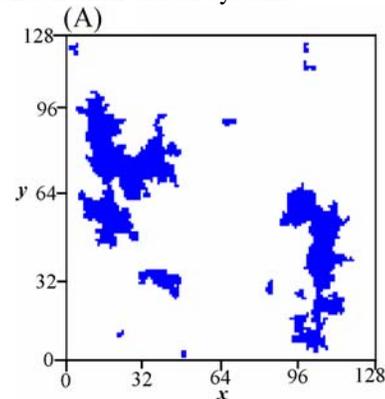

**Figure 1 Clusters that result for $\alpha=1$ and $p=0$.**

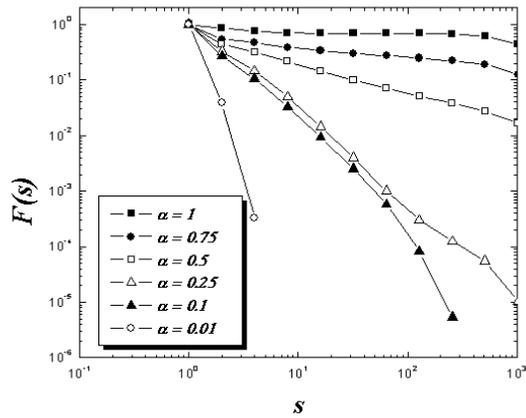

**Figure 2** CSD for different values of α at *p*=0.

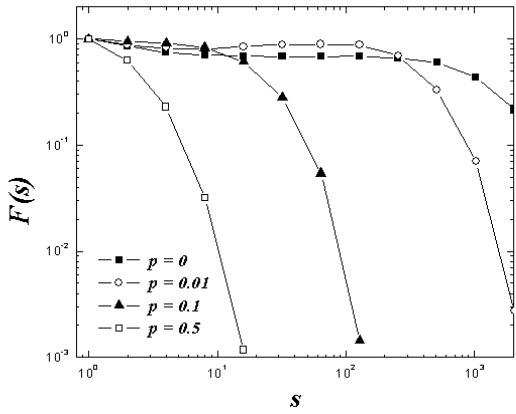

**Figure 3** CSD at α=0 and different values of *p*.

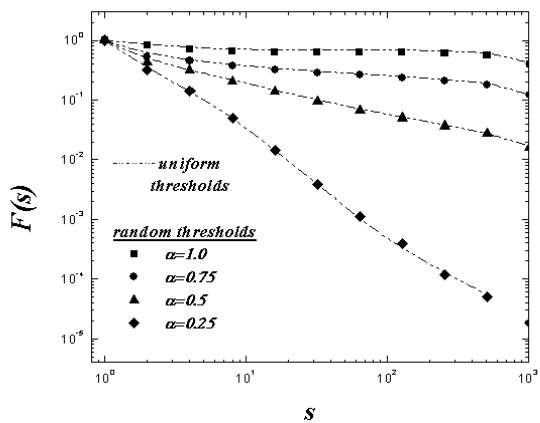

**Figure 4 Comparison between imposing uniform threshold (σ_c=4) and random threshold (2≤σ_c≤16) for different values of α at *p*=0.**

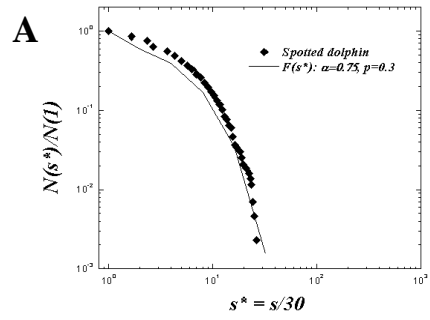

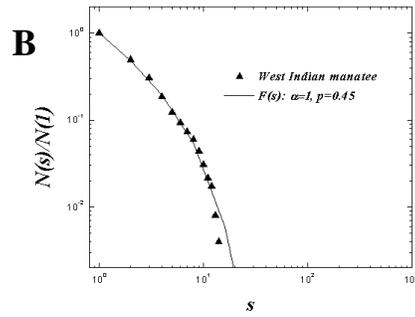

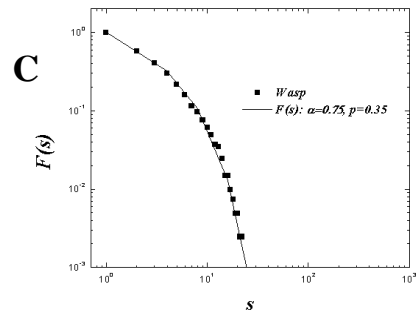

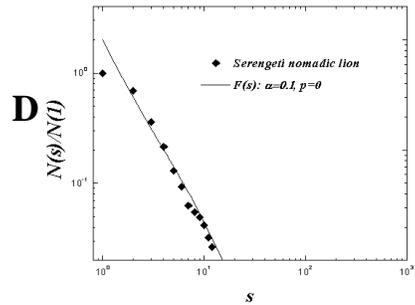

**Figure 5 Fitting the model to four different animal systems. A – Spotted dolphin, *Stenella attenuata* (α=0.75, *p*=0.3, MSE=0.00381); B – West Indian manatee, *Trichecus manatus* (α=1, *p*=0.45, MSE=0.00034); C – Wasp, *Ropalidia fasciata* (α=0.75, *p*=0.35, MSE=0.00059); D – Serengeti lion *Panthera leo* (α=0.1, *p*=0, MSE=0.08661).**

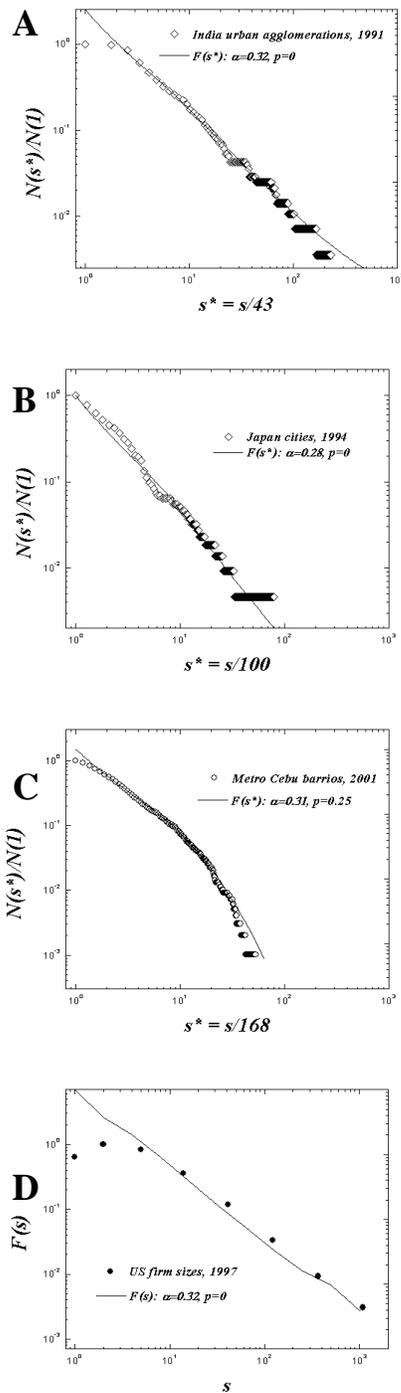

**Figure 6 Fitting the model to four distinct human cluster systems. A – Urban agglomerations of India, 1991 ($\alpha$=0.32, *p*=0, MSE=0.00394); B – Major cities of Japan, 1994 ($\alpha$=28, *p*=0, MSE=0.00021); C – Households/barrios of Metro Cebu, 2001 ($\alpha$=0.31, *p*=0.25, MSE=0.00179); D – Firms/clusters of employees of U.S., 1997 ($\alpha$=0.32, *p*=0, MSE=6.16504).**